\documentclass[aps,prl,twocolumn,superscriptaddress]{revtex4-2}

\usepackage{graphicx,amsmath,amssymb}
\usepackage{hyperref}

\newcommand{\C}{$^{\circ}\mathrm{C}$}
\newcommand{\unit}[1]{\mathord{\mathrm{#1}}}
\newcommand{\ket}[1]{\lvert #1 \rangle}
\begin{document}

\title{Magneto-optical trapping of a group-\uppercase\expandafter{\romannumeral3} atom}
\author{Xianquan Yu}
\affiliation{Centre for Quantum Technologies, National University of Singapore, 3 Science Drive 2, Singapore 117543}
\author{Jinchao Mo}
\affiliation{Department of Physics, National University of Singapore, 2 Science Drive 3, Singapore 117551}
\author{Tiangao Lu}
\affiliation{Centre for Quantum Technologies, National University of Singapore, 3 Science Drive 2, Singapore 117543}
\author{Ting You Tan}
\affiliation{Department of Physics, National University of Singapore, 2 Science Drive 3, Singapore 117551}
\author{Travis L. Nicholson}
\email{nicholson@nus.edu.sg}-
\affiliation{Centre for Quantum Technologies, National University of Singapore, 3 Science Drive 2, Singapore 117543}
\affiliation{Department of Physics, National University of Singapore, 2 Science Drive 3, Singapore 117551}

\begin{abstract}
    We realize the first magneto-optical trap of an atom in main group III of the Periodic Table. Our atom of choice (indium) does not have a transition out of its ground state suitable for laser cooling; therefore, laser cooling is performed on the $\ket{5P_{3/2},F=6} \rightarrow \ket{5D_{5/2},F=7}$ transition, where $\ket{5P_{3/2},F=6}$ is a long-lived metastable state. Optimization of our trap parameters results in atoms numbers as large as $5\times10^8$ atoms with temperatures of order 1 mK. Additionally, through trap decay measurements, we infer a one-body trap lifetime of $12.3\ \unit{s}$. This lifetime is consistent with background gas collisions and indicates that our repumpers have closed all leakage pathways. We also infer a two-body loss rate of $1.6\times 10^{-11}\ \unit{cm^3/s}$, which is comparable to those measured in alkali atoms. The techniques demonstrated in this Letter can be straightforwardly applied to other group-\uppercase\expandafter{\romannumeral3} atoms, and our results pave the way for realizing quantum degenerate gases of these particles.
\end{abstract}

\date{\today}
\maketitle
The magneto-optical trap (MOT) is the workhorse of ultracold physics. This technique has enabled the realization of quantum degenerate gases \cite{Schr2021}, quantum simulators \cite{gross2017,Scholl2021,Ebadi2021}, exotic quantum matter \cite{moses2017}, neutral atom quantum processors \cite{Levine2019,Henriet2020,Kaufman2021,Graham2022}, record-accurate atomic clocks \cite{Ludlow2015}, and precise tests of Standard Model extensions \cite{safronova2018}. Despite the remarkable progress in ultracold physics over the past few decades, most ultracold experiments have been based on alkali metals, alkaline-earth metals, and a small number of lanthanide atoms \cite{Schr2021}. However, the majority of the Periodic Table remains unexplored in the ultracold regime.

One class of atom that has not been cooled to ultracold temperatures is main group \uppercase\expandafter{\romannumeral3} of the Periodic Table, also known as the ``triel elements.'' Triels have many interesting properties that distinguish them from previous ultracold research. Like alkali metals, triels are expected to have ground state magnetic Feshbach resonances \cite{Chin2010}, which have not been found in gases of only alkaline earths. Magnetic Feshbach resonances allow for precise control over quantum many-body states, and they have been the focus of many years of impactful ultracold experiments. However, like alkaline earths, the triel atoms thallium and indium have narrow linewidth electronic transitions that are amenable to stable laser technology \cite{footnote1}. These transitions (which are not found in alkalis) are used in highly accurate atomic clocks \cite{Ludlow2015}, and they also allow for precise probes of atoms' internal states and their interactions \cite{martin2013,zhang2014,scazza2014}. Therefore, triels could be probed with the precision of an atomic clock while offering the many-body control of alkali metals. Furthermore, unlike the $S$-orbital ground states of alkali metals and alkaline earths or the high angular momentum ground states of popular lanthanides (such as erbium and dysprosium), the $P$-orbital ground states of triels distinguish themselves as intermediate cases. Low-temperature scattering between such states has not been studied, and these interactions could prove as interesting and surprising as scattering measurements with lanthanides \cite{Petrov2012,Maier2015,Khlebnikov2019,Norcia2021}.

The main challenge of realizing a triel MOT is that these atoms do not contain cycling transitions amenable to laser cooling out of their $P_{1/2}$ ground states. Earlier attempts to transversely cool beams of the triel atom indium concluded that cooling on the $5P_{1/2} \rightarrow 6S_{1/2}$ transition is inefficient even with repumping \cite{Kloter2008}. However, the long lived metastable $P_{3/2}$ states in triel atoms offer a $P_{3/2} \rightarrow D_{5/2}$ cycling transition that is suitable for laser cooling. Using this transition, transverse cooling was observed in the triel atoms Al \cite{McGowan95}, Ga \cite{Rehse2004}, and In \cite{Kim2009}, and our team recently realized an indium Zeeman slower \cite{Yu2022}. Laser cooling of the triel Tl has also been proposed with this transition \cite{Fan2011}. Although most MOTs are formed using transitions out of atomic ground states, a MOT based on a transition out of an optically pumped metastable electronic state (in the lanthanide europium) \cite{Inoue2018} has been demonstrated. However, a MOT has never been realized with a triel atom.

\begin{figure*}[htbp]
	\centering
	\includegraphics[width=\linewidth]{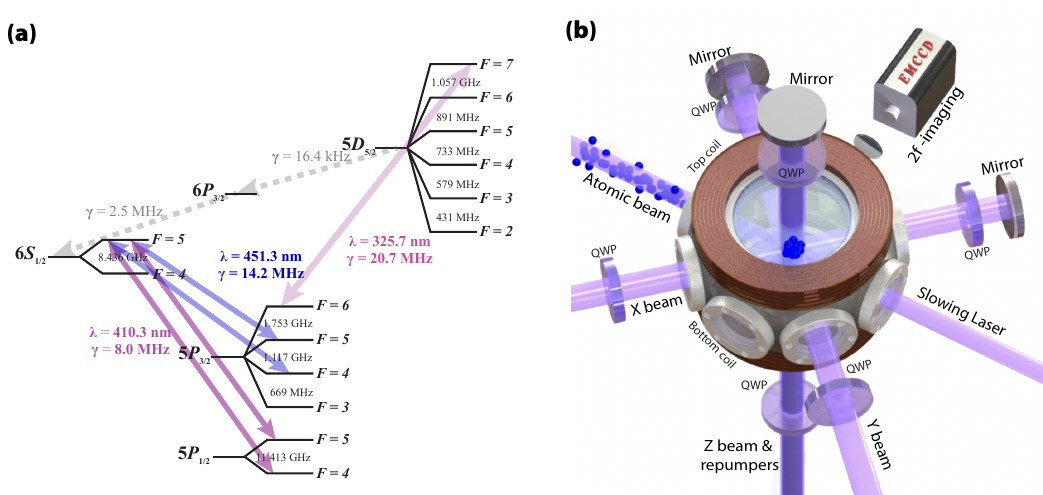}
	\caption{(a) Energy levels of $^{115}\mathrm{In}$ \cite{Kim2009,Eck1957,Gunawardena2009,Zimmermann1970,Safr2007}. Relevant transition wavelengths and natural linewidths are denoted with $\lambda$ and $\gamma = \Gamma/2\pi$, respectively. Laser cooling of indium is based on the cycling transition $\ket{5P_{3/2},F=6} \rightarrow \ket{5D_{5/2},F=7}$, where the lower-energy cooling state is long lived ($\sim10\ \unit{s}$ lifetime) \cite{Sahoo2011}. We drive the two $\ket{5P_{1/2},F=4,5} \rightarrow \ket{6S_{1/2},F=5}$ transitions at 410~nm and the two $\ket{5P_{3/2},F=4,5} \rightarrow \ket{6S_{1/2},F=5}$ transitions at 451~nm for initial state preparation just after the atomic beam emerges from the effusion cell \cite{Yu2022}. Lasers on these four transitions are also used for repumping during the Zeeman slowing and MOT stages. (b) The magneto-optical trap setup. Three pairs of orthogonal 326~nm cooling beams in $\sigma^+ - \sigma^-$ configuration are sent through the center of the vacuum chamber, which overlaps with the center of the quadrupole magnetic field generated by a pair of coils in anti-Helmholtz configuration. Two 410~nm and two 451~nm repumper lasers are coaligned with the $z$ direction MOT beam. The MOT fluorescence signals are collected by an electron multiplying CCD (EMCCD) camera via a $2f$ imaging system.}
	\label{fig:levels_MOT}
\end{figure*}

In this Letter, we demonstrate the first MOT of a group-\uppercase\expandafter{\romannumeral3} atom using $^{115}\mathrm{In}$. This species is a boson with a nuclear spin of $I=9/2$ and an isotopic abundance of 95.7\%. The $^{115}\mathrm{In}$ energy level diagram and relevant transitions for this work are shown in Fig. \ref{fig:levels_MOT}(a). For efficient laser cooling, we use the 326~nm $\ket{5P_{3/2},F=6} \rightarrow \ket{5D_{5/2},F=7}$ transition. Possible mechanisms that cause population to leak out of this transition's states are the $5P_{3/2} \rightarrow 5P_{1/2}$ (lifetime predicted to be $\sim 10\ \unit{s}$ \cite{Sahoo2011}), the $5D_{5/2} \rightarrow 6P_{3/2} \rightarrow 6S_{1/2}$ decay pathway (which occurs with a branching ratio of order $10^{-4}$ \cite{Safr2007}), and off-resonant driving of the $\ket{5P_{3/2},F=6} \rightarrow \ket{5D_{5/2},F=6}$ transition. To close these leaks, we repump population into the $\ket{5P_{3/2},F=6}$ cooling state with two 410~nm repumpers (which address the $\ket{5P_{1/2},F=4,5} \rightarrow \ket{6S_{1/2},F=5}$ transitions) and two 451~nm repumpers (which address the $\ket{5P_{3/2},F=4,5} \rightarrow \ket{6S_{1/2},F=5}$ transitions).

An indium atomic beam is produced by an effusion cell operating at 800 \C, resulting in a $~10^{-3}\ \unit{Torr}$ vapor pressure inside the crucible. The output of the effusion cell is collimated with 200 microchannels that are 1 cm in length and $200\ \unit{\mu m}$ in diameter. Atoms that emerge from the cell are pumped out of the $5P_{1/2}$ ground state with a pair of 410~nm lasers, which results in ample population in the $\ket{5P_{3/2},F=6}$ cooling state due to a favorable branching ratio. Additionally, we use two 451~nm lasers to pump the remaining population that decays into $\ket{5P_{3/2},F=4,5}$ into $\ket{5P_{3/2},F=6}$. The atomic beam is then slowed to a $70\ \unit{m/s}$ velocity with a transverse field permanent magnet Zeeman slower \cite{Yu2022}.

The slowed atomic beam enters a vacuum chamber with a pressure at the low $10^{-10}\ \unit{Torr}$ level. A pair of coils in anti-Helmholtz configuration generate a quadrupole magnetic field that has an axial field gradient up to $55\ \unit{G/cm}$ at the center of the chamber [Fig. \ref{fig:levels_MOT}(b)]. Three pairs of orthogonal counter-propagating 325.7 nm cooling beams, which have a $1/e^2$ diameter of 12~mm, are circularly polarized in the traditional $\sigma_+ - \sigma_-$ configuration. In addition, the four repumpers, which also have a $1/e^2$ diameter of 12~mm, are coaligned with the $z$-direction MOT beam. The cooling laser is generated by frequency quadrupling the output of a 1302.8~nm external-cavity-diode-laser-seeded Raman fiber amplifier, resulting in hundreds of milliwatts of useful 325.7~nm laser power. The 325.7~nm laser is frequency stabilized to a calibrated, low-drift HighFinesse wavemeter. Meanwhile the four repumper beams are generated by external cavity diode lasers stabilized to indium hollow cathode lamps with Doppler-free spectroscopy.

With the cooling beams and all four repumpers in place, we observe a bright MOT signal. MOT fluorescence is collected with a $2f$ imaging system focused onto an EMCCD camera. The number of atoms $N$ in the MOT is inferred from a fluorescence image as \cite{supplemental}

\begin{equation}
\label{eq:atom_number}
    N = \frac{8\pi N_c}{\eta\Omega\Gamma t_{ex} s_0} \left(1+s_0 + \frac{4\Delta^2}{\Gamma^2} \right),
\end{equation}
where $N_c$ is the total number of fluorescence counts in a MOT image, $\Omega$ is the solid angle subtended by the imaging system, $\eta$ is the imaging system's quantum efficiency, $t_{ex}$ is the exposure time of the camera, $\Delta$ is the MOT detuning (in rad/s), and $\Gamma$ is the natural linewidth (in rad/s) of the cooling transition. The saturation parameter of the MOT lasers is $s_0 = I/\mathcal{M} I_{sat}$, where $I$ is the total intensity of all six MOT laser beams and $I_{sat} = 78.3\ \unit{mW/cm^2}$ is the saturation intensity of the cooling transition. The multiplicity factor $\mathcal{M} = 3(2F+1)/(2F+3) = 2.6$ (where $F$ is associated with the lower-energy cooling state) accounts for the fact that the cooling transition is not a two-level system; rather, the MOT acts on the Zeeman sublevels of each cooling state, and population is assumed to be distributed roughly evenly among these sublevels \cite{supplemental,Bradley2000,McClelland2006,Lu2010,Hostetter2014,Inoue2018,Java1993,muller2005}.

Although Eq. (\ref{eq:atom_number}) ignores the magnetic field, the field's fractional contribution to the atom number calculation is  \cite{supplemental}

\begin{equation}
    \frac{\Delta N}{N} = \left( \frac{2 \mu}{\Gamma} \frac{\partial B}{\partial z} \right)^2 \frac{\sigma_z^2 + \frac12 \sigma_x^2}{1 + s_0 + \frac{4\Delta^2}{\Gamma^2}}.
\end{equation}
Here $\Delta N$ is the change in the computed atom number due to the magnetic field, $\mu$ is the effective magnetic moment, $\partial B/\partial z$ is the $z$ derivative of the magnetic field evaluated at the trap center (taken to be the origin), and $\sigma_{x,y,z}$ are the rms radii of the Gaussian atomic density distribution. We estimate that including the magnetic field results in a 4\% correction to Eq. (\ref{eq:atom_number}), so we neglect this.

\begin{figure}[htbp]
	\centering
	\includegraphics[width=\linewidth]{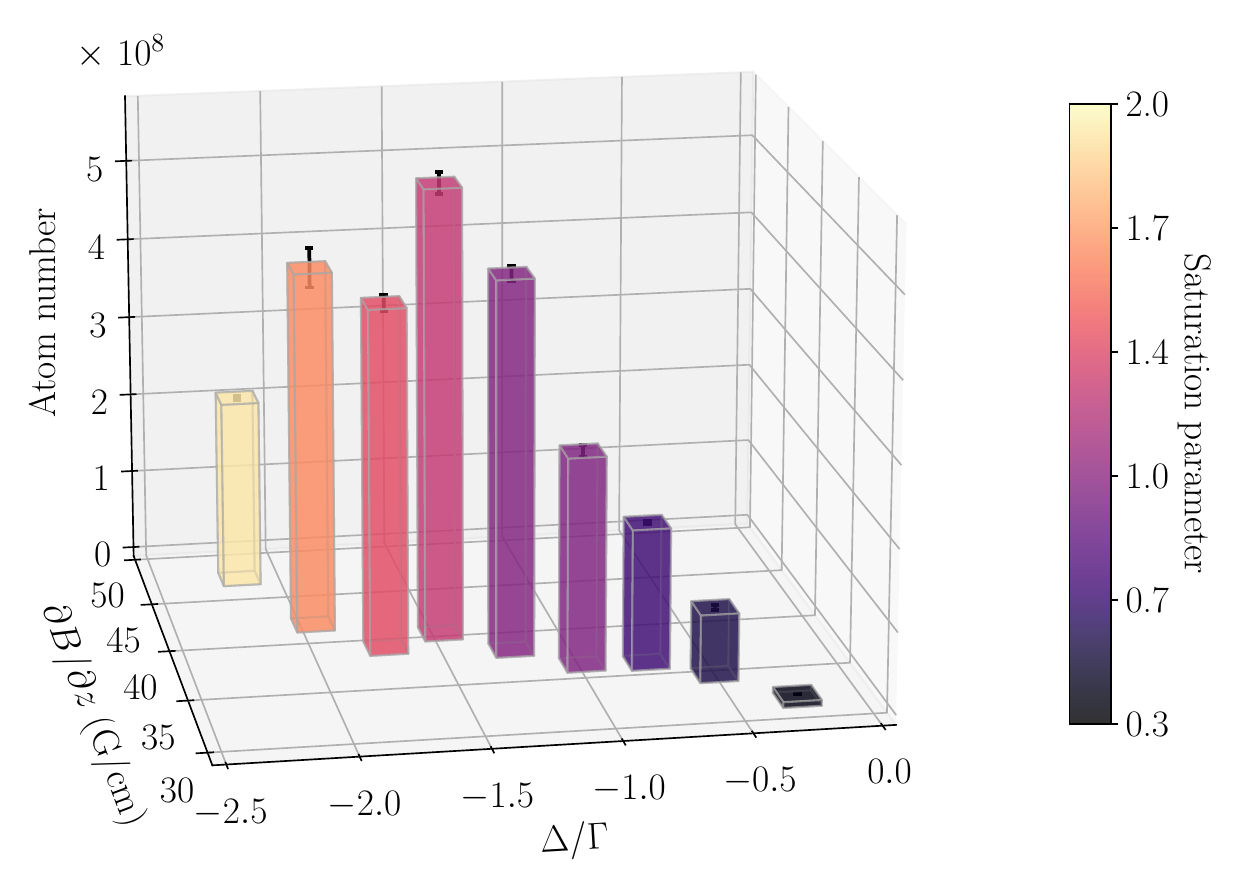}
	\caption{Optimization of the MOT atom number. We vary the MOT detuning $\Delta$, the MOT laser intensity (expressed in the figure as the saturation parameter $s_0 = I/\mathcal{M}I_{sat}$), and the magnetic field gradient $\partial B/\partial z$ at the trap center projected along the $z$ axis. The measurement is taken by fixing the detuning and then varying $s_0$ and $\partial B/\partial z$ until the atom number maximizes. Each bar represents the best atom number observed for a given detuning. The trapped atom number fluctuations are estimated from the standard deviation of multiple atom number measurements for the same MOT parameters.}
	\label{fig:optimization}
\end{figure}

To optimize the total atom number, we systematically vary $\Delta$. Once a value of $\Delta$ is fixed, we vary the per-beam MOT laser power $P$ and $\partial B/\partial z$ until the atom number is maximized. We find a global maximum of the atom number when $\Delta/\Gamma = -1.6$, $\partial B/\partial z = 36\  \unit{G/cm}$, and $P = 24.5\ \unit{mW}$, the latter of which corresponds to $s_0 = 1.3$ (Fig. \ref{fig:optimization}). With these values, we observe a MOT with $5\times10^8$ atoms. The uncertainty due to random fluctuations in $N$ is obtained by measuring the atom number several times for a fixed set of MOT parameters. The standard deviation in these measurements divided by their mean is treated as the fractional uncertainty in $N$ and applied to the remainder of the optimization data (Fig. \ref{fig:optimization}).

With the MOT parameters fixed at their optimum values, we measure the temperature of the trapped gas with a time-of-flight measurement \cite{supplemental}. We observe the vertical (horizontal) temperature to be $4.5\ \unit{mK}$ ($3.6\ \unit{mK}$) for optimal MOT parameters, whereas the predicted temperature from Doppler theory is $1\ \unit{mK}$. MOT temperatures above the Doppler value have been observed in multiple experiments \cite{chang2014,Zinner2000,Sengstock1994,Riedmann2012,Xu2002,Kuwamoto1999,Loo2003}.

An important characterization of a MOT is its one- and two-body loss rates. A small one-body lifetime can indicate that our repumping scheme has not closed all population leaks out of the cooling states. This is especially a concern since our cooling is based on a metastable state, which results in many more levels lower in energy than the cooling transition (and therefore more decay pathways) compared to alkali-metal and alkaline-earth-metal MOTs that cool out of a ground state hyperfine manifold. In fact, another realization of a MOT using an optically pumped metastable electronic transition found a small one-body lifetime and identified uncontrolled population leaks as the cause \cite{Inoue2018}. Additionally, a measurement of the two-body rate can determine the severity of unwanted effects, such as light-assisted collisions.

\begin{figure}[htbp]
	\centering
	\includegraphics[width=\linewidth]{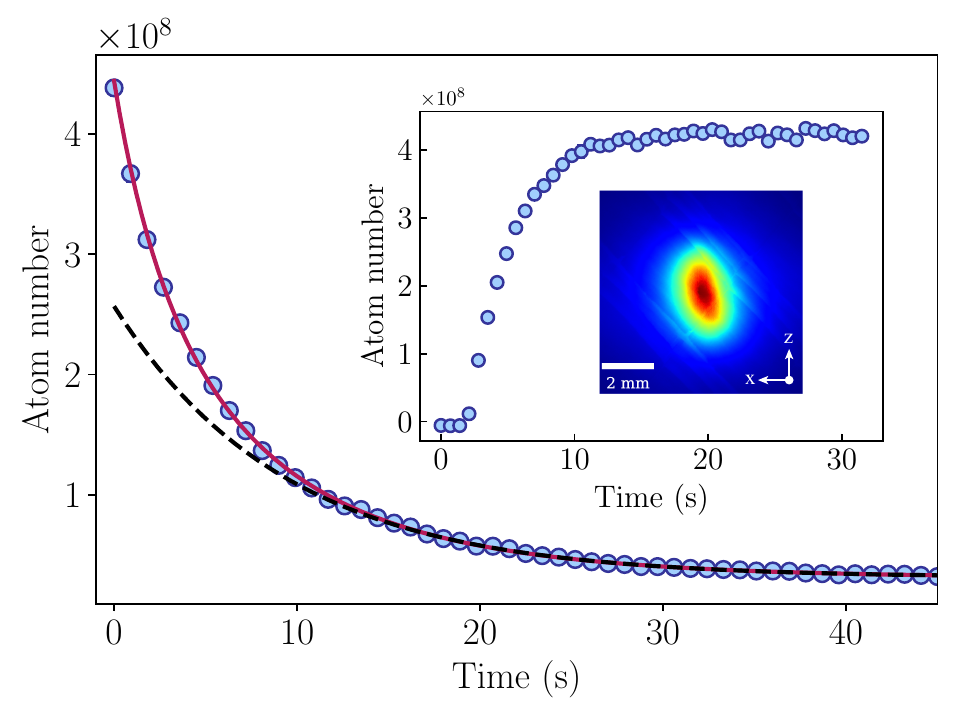}
	\caption{Atom number decay for optimal MOT parameters. The red solid curve is generated by fitting Eq. (\ref{eq:decay}) to the data. The black dashed curve is the solution to Eq. (\ref{eq: differential equation}) with $\beta$ taken to be zero and fit to the data for times larger than $15\ \unit{s}$. The $\beta = 0$ case is a pure one-body model, which is expected to be valid at longer times after two-body processes have decayed. The discrepancy between the data and the black dashed line at short times illustrates the extent of two-body effects. The inset: Observed MOT loading curve and a false color MOT fluorescence image.  The fitted loading rate for these data is $1.30(2) \times 10^8\ \unit{atoms/s}$. A two-dimensional Gaussian fit of the steady-state MOT fluorescence image yields root-mean-square radii $\sigma_x = 1.0\ \unit{mm}$ and $\sigma_z = 1.3\ \unit{mm}$.}
	\label{fig:long_decay}
\end{figure}

We measure these rates by disrupting MOT loading and observing the resulting atom number decay (Fig. \ref{fig:long_decay}). The MOT loading rate is changed by shutting off the Zeeman slower laser, which results in a greatly reduced steady-state MOT number. Our decay data is taken with the optimal MOT parameters (found above) as well as $7\ \unit{mW}$ of combined 410-nm laser repumper power and $7\ \unit{mW}$ of combined 451-nm laser repumper power.

The decay of the trapped atom number $N(t)$ in the MOT can be modeled by a rate equation,
\begin{equation}
    \frac{dN}{dt} = L - \frac{N}{\tau} - \beta\int n^2(\vec{r},t)d^3r\ ,
    \label{eq: differential equation}
\end{equation}
where $L$ is the MOT loading rate (in this case, the rate without the Zeeman slower laser), $\tau$ is the MOT one-body lifetime, and $\beta$ is the two-body rate constant. When the MOT has a Gaussian atomic density distribution, we can simplify Eq. (\ref{eq: differential equation}) by rewriting the two-body loss term as $-\Gamma_2 N^2$, where $\Gamma_2 = \beta/(8\pi^{3/2}\sigma_x\sigma_y\sigma_z)$. The decay of the trapped atom number $N(t)$ is therefore given by

\begin{equation}
    N(t) = \frac{2L\tau-(1-\kappa)N_0-[2L\tau-(1+\kappa)N_0]e^{-\kappa t/\tau}}{1+\kappa+2N_0\Gamma_2\tau-(1-\kappa+2N_0\Gamma_2\tau)e^{-\kappa t/\tau}}\ ,
    \label{eq:decay}
\end{equation}
where $\kappa = \sqrt{1+4L\Gamma_2\tau^2}$ and $N_0$ is the atom number at $t = 0$. The MOT lifetime and two-body loss rate can be extracted from fitting the decay data to Eq. (\ref{eq:decay}). For the one-body lifetime, we obtain a value of $\tau = 12.3(3)\ \unit{s}$. This number is consistent with decay due to collisions with the background gas, which is the technical limit of the MOT lifetime. Meanwhile, the fitted two-body rate constant is $\beta = 1.6(5)\times10^{-11}\ \unit{cm^3/s}$. We have observed that the fitted value of $\beta$ varies by 30\% between different runs of our experiment, and this variation is reflected in our quoted uncertainty for $\beta$. For comparison, two-body rate constants measured in well-functioning MOTs of alkali-metal atoms are found to be $4\times10^{-11}\ \unit{cm^3/s}$ for $^{23}\mathrm{Na}$ \cite{Marcassa1993}, $5.8\times10^{-12}\ \unit{cm^3/s}$ for $^{87}\mathrm{Rb}$ \cite{Wallace1992}, and $7.6\times10^{-11}\ \unit{cm^3/s}$ for $^{133}\mathrm{Cs}$ \cite{Sesko1989}.

\begin{figure}[htbp]
	\centering
	\includegraphics[width=\linewidth]{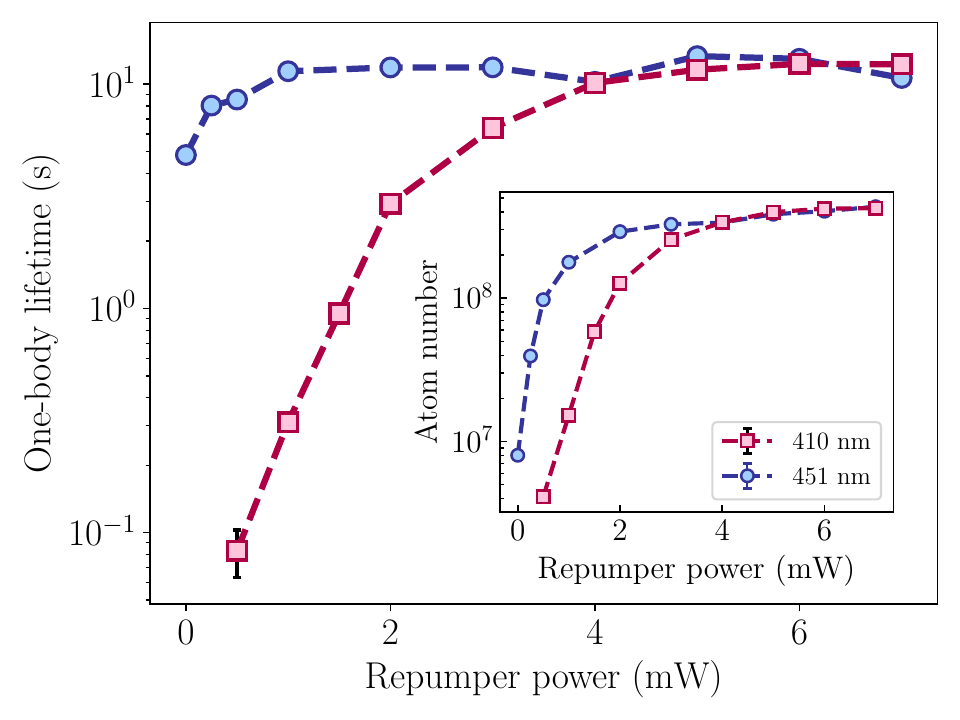}
	\caption{MOT lifetime as a function of repumper power. These data are collected under the optimal MOT parameters mentioned above. The $x$ axis is the combined power of both repumpers for a given wavelength. With the repumper lasers, the MOT lifetime extends up to two orders of magnitude. The inset: Trapped atom number as a function of repumper power under optimal MOT parameters. The trapped atom number increases up to two orders of magnitude with the application of repumpers.}
	\label{fig:repumpers}
\end{figure}

To further study repumping efficiency, we vary repumper powers and measure the effect on the MOT one-body lifetime and steady-state atom number. Data are collected by varying the combined power of both 410-nm (451-nm) repumpers whereas the power of the 451-nm (410-nm) repumpers is fixed at its maximum value. Below 4 mW (1 mW) of 410-nm (451-nm) laser light, we observe that the one-body lifetime and steady-state atom number depends strongly on repumper power (Fig. \ref{fig:repumpers}). Above this regime, the repumping transitions are saturated, and the MOT has a large atom number and a long one-body lifetime. Comparing the 410- and 451-nm repumping data, it is clear that 451-nm repumping saturates more easily. This is because the 410-nm lasers are somewhat effective at populating $\ket{5P_{3/2},F=6}$ on their own due to a 60\% branching ratio for decay into this state.

We compare this result to MOT realizations that are reportedly limited by repumper power or population leakage pathways that are not fully closed \cite{Inoue2018,Hostetter2014,Sukachev2010,De2009,Gr2001,Mills2017}. In these cases, one-body lifetimes are an order of magnitude shorter than in our apparatus. The data reported here suggest that population leaks are well closed by our repumpers.

In conclusion, we demonstrate a magneto-optical trap of a group-\uppercase\expandafter{\romannumeral3} atom. Over $5\times10^8$ atoms are loaded to the MOT from a Zeeman slowed atomic beam. With ample repumper power, measurements of MOT decay implies a one-body lifetime of $\tau = 12.3(3)\ \unit{s}$. This lifetime as well as repumper characterization confirms that all leakage pathways are sufficiently closed. Meanwhile, we observed a two-body rate of $\beta = 1.6(5) \times10^{-11}\ \unit{cm^3/s}$. Further improvements are possible by increasing the atomic density using dark spot MOT or compressed MOT techniques, and lower temperatures can be achieved with Sisyphus cooling. Techniques demonstrated in this Letter can be extended to all atoms with similar atomic structure, such as the group-\uppercase\expandafter{\romannumeral3} atoms Tl, Al, Ga, and B. Furthermore, our Letter paves the way of exploring group-\uppercase\expandafter{\romannumeral3} atoms in the quantum degenerate regime.

\begin{acknowledgments}
This work was supported by the National Research Foundation, Prime Minister’s Office, Singapore and the Ministry of Education, Singapore under the Research Centres of Excellence program.
\end{acknowledgments}

\bibliography{references.bib}

\end{document}


\title{Magneto-optical trap of a Group III atom: Supplemental material}
\author{Xianquan Yu}
\affiliation{Centre for Quantum Technologies, National University of Singapore, 3 Science Drive 2, Singapore 117543}
\author{Jinchao Mo}
\affiliation{Department of Physics, National University of Singapore, 2 Science Drive 3, Singapore 117551}
\author{Tiangao Lu}
\affiliation{Centre for Quantum Technologies, National University of Singapore, 3 Science Drive 2, Singapore 117543}
\author{Ting You Tan}
\affiliation{Department of Physics, National University of Singapore, 2 Science Drive 3, Singapore 117551}
\author{Travis L. Nicholson}
\affiliation{Centre for Quantum Technologies, National University of Singapore, 3 Science Drive 2, Singapore 117543}
\affiliation{Department of Physics, National University of Singapore, 2 Science Drive 3, Singapore 117551}

\date{\today}
\maketitle

\section{Inferring the atom number from a MOT fluorescence measurement}
\label{sec:mot_number}
We consider a small volume of space $dV$ located at coordinates $x$, $y$, and $z$. The volume is small enough that the external magnetic field gradient does not vary significantly within it and can therefore be taken as constant. The rate of photons emitted from this volume is

\begin{equation}
    d\Gamma_{ph} = \Gamma_{sc} dN,
\end{equation}
where $dN$ is the atom number within $dV$, and the single-atom scattering rate $\Gamma_{sc}$ is

\begin{equation}
    \Gamma_{sc} = \frac{\Gamma}{2} \frac{s_0}{1 + s_0 + \frac{4}{\Gamma^2} (\Delta + \mu B)^2}.
\end{equation}
Here $\Gamma$ is the spontaneous emission rate from the MOT transition, $s_0$ is the saturation parameter, $\Delta$ is the MOT detuning, and $B$ is the MOT's external magnetic field. The effective magnetic moment of the MOT transition is $\mu = (g_e m_e - g_g m_g)\mu_B$, where $g_e$ ($g_g$) is the excited (ground) state $g$-factor, $m_e$ ($m_g$) is the excited (ground) state magnetic quantum number, and $\mu_B$ is the Bohr magneton. 

The quadrupole field $B$ can be approximated near the origin as

\begin{equation}
    B \simeq \beta_x x + \beta_y y + \beta_z z.
\end{equation}
Here $\beta_i$ is defined as

\begin{equation}
    \beta_i = \frac{\partial B}{\partial i}\bigg|_{x=y=z=0},
\end{equation}
with $i=x,y,z$. 

The rate of photons emitted from $dV$ that make it to the camera is
\begin{equation}
    d\Gamma_{cam} = \frac{\Omega}{4\pi} d\Gamma_{ph} = \frac{\Omega}{4\pi} \Gamma_{sc} dN,
\end{equation}
where $\Omega$ is the solid angle subtended by the imaging system. One useful expression for this analysis is the density of camera counts $n_c$, which can be obtained by fitting a picture of the MOT fluorescence with a Gaussian, resulting in

\begin{equation}
    n_c = \frac{N_c}{(2\pi)^{3/2}\sigma_x \sigma_y \sigma_z} e^{-x^2/2 \sigma_x^2} e^{-y^2/2 \sigma_y^2} e^{-z^2/2 \sigma_z^2}.
\end{equation}
Here $N_c$ is the total integrated camera counts, and $\sigma_x$, $\sigma_y$, and $\sigma_z$ are the fitted RMS widths of the MOT (usually only two of these widths can be fit and the other is inferred by symmetry). Using this expression, the number of camera counts due to the fluorescence originating from $dV$ is

\begin{equation}
    n_c dV = \frac{\Omega}{4\pi} \Gamma_{sc} \eta t_{ex} dN,
\end{equation}
where $\eta$ is the camera's quantum efficiency and $t_{ex}$ is its exposure time. Or in terms of densities,

\begin{equation}
    n_c = \frac{\Omega}{4\pi} \Gamma_{sc} \eta t_{ex} n,
\end{equation}
where $n$ is the atom number density.

The total atom number is therefore

\begin{align}
    N &= \int n \, dV \\
    &= \frac{4\pi}{\Omega \eta t_{ex}} \int \frac{n_c}{\Gamma_{sc}} dV \\
    &= \frac{8\pi}{\Omega\Gamma \eta t_{ex}s_0} \left[ \left(1+s_0 + \frac{4\Delta^2}{\Gamma^2} \right)N_c + \frac{4}{\Gamma^2} \int (2\Delta\mu B + \mu^2 B^2) n_c dV \right].
    \label{eqn:number1}
\end{align}
Note since $B$ is an odd function in its spatial coordinates whereas $n_c$ is an even one, the first term in the integral of Eqn. \ref{eqn:number1} vanishes. The main text (and several other references) estimates the atom number by ignoring the magnetic field, resulting in the approximate number expression

\begin{equation}
    N_{approx} = \frac{8\pi N_c}{\Omega\Gamma \eta t_{ex}s_0} \left(1+s_0 + \frac{4\Delta^2}{\Gamma^2} \right).
\end{equation}
The correction to $N_{approx}$ due to the magnetic field is therefore

\begin{align}
    N - N_{approx} &= \frac{32\pi\mu^2}{\Omega\Gamma^3 \eta t_{ex} s_0} \int B^2 n_c dV \\
    &= \frac{32\pi\mu^2 N_c}{\Omega\Gamma^3 \eta t_{ex} s_0} \left(\beta_x^2 \sigma_x^2 + \beta_y^2 \sigma_y^2 + \beta_z^2 \sigma_z^2 \right).
\end{align}
Using the fact that $\sigma_x = \sigma_y$ and $\beta_x = \beta_y$ (due to the symmetry of the anti-Helmholtz field) as well as $\beta_x + \beta_y + \beta_z = 0$ (Gauss' law for magnetism), we can write the fractional correction to the approximate atom number calculation as

\begin{equation}
    \frac{N-N_{approx}}{N_{approx}} = \frac{4\mu^2\beta_z^2}{\Gamma^2} \frac{\sigma_z^2 + \frac12 \sigma_x^2}{1 + s_0 + \frac{4\Delta^2}{\Gamma^2}}.
\end{equation}

For the optimal numbers found in the main text, $\mu = \mu_B$, $\beta_z = 36\ \unit{G/cm}$, $s_0 = 1.3$, and $\Delta/\Gamma = -1.6$. Also, $\sigma_x = 1.3\ \unit{mm}$ and $\sigma_z = 1\ \unit{mm}$. This results in a correction of 3.5\% to the atom numbers quoted in the main text. All numbers in the main text neglect this correction.

\section{Saturation intensity multiplicity factor}
Consider an atomic transition driven by a laser field. The saturation intensity is

\begin{equation}
    I_{sat} = \frac{\hbar\omega}{2\sigma\tau},
\end{equation}
where $\omega$ is the laser frequency, $\sigma$ is the photon scattering cross section, and $\tau$ is the transition lifetime. For a two-level system, the photon scattering cross section is

\begin{equation}
\label{eq:two_level_cross_section}
    \sigma = \frac{3\lambda^2}{2\pi} \frac{1}{1 + \frac{4\Delta^2}{\Gamma^2}}.
\end{equation}
Here, $\lambda$ is the photon wavelength. On resonance, this results in the familiar expression for the saturation intensity

\begin{equation}
\label{eq:two_level_saturation_intensity}
    I_{sat} = \frac{\pi h c}{2 \lambda^3\tau}.
\end{equation}

When the laser is driving a transition between two atomic hyperfine states, each with a manifold of Zeeman sublevels, Eq. \ref{eq:two_level_cross_section} is no longer valid. However, it can be replaced with the effective cross section \cite{muller2005}

\begin{align}
    \sigma_{eff} &= \frac{3\lambda^2}{2\pi} \frac{\mathcal{M}}{1 + \frac{4\Delta^2}{\Gamma^2}} \\
    &= \mathcal{M} \sigma,
\end{align}
where the multiplicity factor $\mathcal{M}$ is

\begin{equation}
    \mathcal{M} = \frac{3(2 F + 1)}{2 F + 3}.
\end{equation}
Therefore, in the presence of Zeeman sublevels, the resonant saturation intensity is $\mathcal{M} I_{sat}$, where $I_{sat}$ is the two-level saturation intensity of Eq. \ref{eq:two_level_saturation_intensity}.

\section{Time-of-flight measurement of MOT temperature}
\begin{figure}[htbp]
	\centering
	\includegraphics[width=0.5\linewidth]{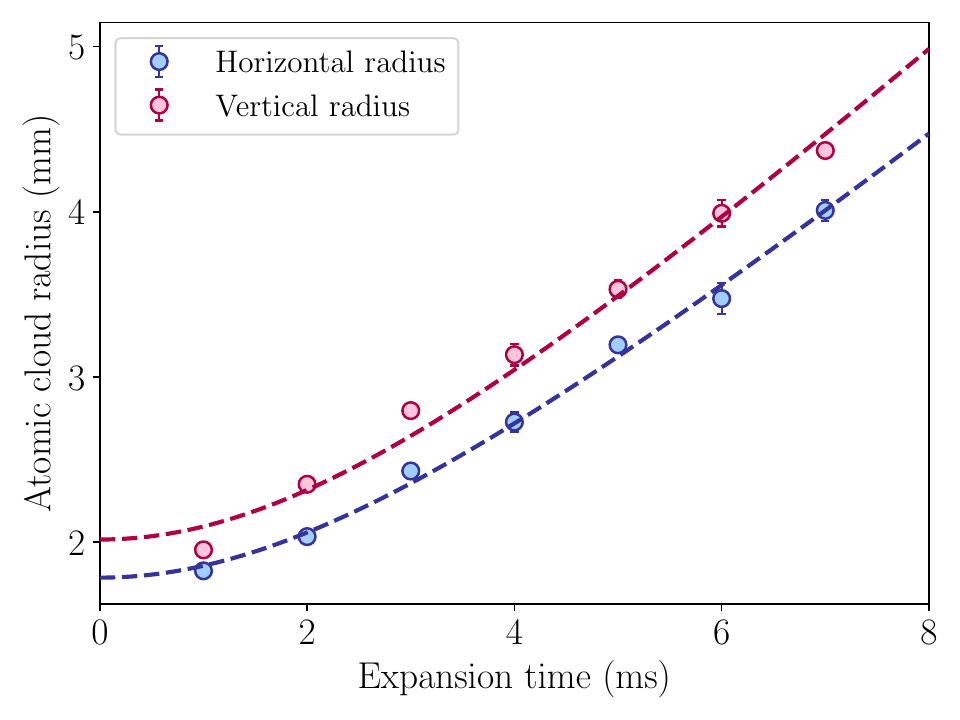}
	\caption{The atomic cloud radius after a variable expansion time. The MOT is generated with the optimal trapping parameters quoted in the main text.}
	\label{fig:tof_measurement}
\end{figure}

The temperature of the MOT is inferred with a time-of-flight measurement. Here the MOT is filled to its steady-state number and then the cooling lasers and magnetic field are abruptly shut off. After a delay (known as the ``expansion time''), the cooling lasers are pulsed back on for $200\ \unit{\mu s}$. During this pulse, the EMCCD camera captures an image of the expanding, falling cloud. 

To analyze the image, we model the atomic cloud with a Gaussian spatial distribution. The root-mean-square radius of the distribution is $\sigma_i(t)$, where $i = x,y,z$. The time evolution of the expanding atomic cloud radius is

\begin{equation}
    \sigma_i(t)^2 = \sigma_i(0)^2 + \frac{k_B T_i}{m}t^2,
    \label{eq:tof_expansion}
\end{equation}
where $\sigma_i(0)$ is the initial cloud radii, $k_B$ is the Boltzmann constant, $m$ is the mass of the $^{115}\mathrm{In}$ atom, $t$ is the expansion time, and $T_i$ is atomic cloud temperature along $i$ axis. The temperature is inferred by varying the expansion time and fitting measured values of $\sigma_i(t)$ to Eq. \ref{eq:tof_expansion}, using $\sigma_i(0)$ and $T_i$ as free parameters.

We obtain the horizontal MOT temperature to be 3.6(1) mK and vertical MOT temperature to be 4.5(1) mK.

\bibliography{supplemental_references.bib}